\documentclass[aps,showpacs,pra,superscriptaddress,]{revtex4}
\usepackage{amsmath}
\usepackage{amsfonts}
\usepackage{amssymb}
\usepackage{bm}
\usepackage{graphicx}

\begin{document}

\title{Chirped periodic and localized waves in a weakly nonlocal media with cubic-quintic nonlinearity}
\author{Houria Triki}
\affiliation{Radiation Physics Laboratory, Department of Physics, Faculty of Sciences,
Badji Mokhtar University, P. O. Box 12, 23000 Annaba, Algeria }
\author{ Vladimir I. Kruglov}
\affiliation{Centre for Engineering Quantum Systems, School of Mathematics and Physics,
The University of Queensland, Brisbane, Queensland 4072, Australia}

\begin{abstract}
We study the propagation of one-dimentional optical beams in a weakly
nonlocal medium exhibiting cubic-quintic nonlinearity. A nonlinear equation
governing the evolution of the beam intensity in the nonlocal medium is
derived thereby which allows us to examine whether the traveling-waves exist
in such optical material. An efficient transformation is applied to obtain
explicit solutions of the envelope model equation in the presence of all
material parameters. We find that a variety of periodic waves accompanied
with a nonlinear chirp do exist in the system in the presence of the weak
nonlocality. Chirped localized intensity dips on a continuous-wave
background as well as solitary waves of the bright and dark types are
obtained in a long wave limit. A class of propagating chirped self-similar
solitary beams is also identified in the material with the consideration of
the inhomogeneities of media. The applications of the obtained self-similar
structures are discussed by considering a periodic distributed amplification
system.
\end{abstract}

\maketitle
\affiliation{$^{1}${\small Radiation Physics Laboratory, Department of Physics, Faculty
of Sciences, Badji Mokhtar University, P. O. Box 12, 23000 Annaba, Algeria}\\
$^{2}${\small Centre for Engineering Quantum Systems, School of Mathematics
and Physics, The University of Queensland, Brisbane, Queensland 4072,
Australia}}

\section{Introduction}

Propagations of spatial optical solitons through nonlocal nonlinear media
have drawn considerable attention because of their experimental observations
in liquid crystals \cite{Conti} and lead glasses \cite{R1} in addition to
several important theoretical predictions \cite{K1,G1,G2}. Nonlocal
nonlinearity represents the fact that the refractive index change of a
material at a particular location, is determined by the light intensity in a
certain vicinity of this location \cite{K1}. Compared with the local
nonlinear medium for which the response at a given point is only dependent
on the light intensity at that point \cite{Kro}, the response of the
nonlocal nonlinear medium at a given point depends not only on the optical
intensity at that point, but also on the intensity in its vicinity \cite%
{Kro,Jia,W1}. It is noteworthy that the nonlocality plays a vital role for
the very narrow beam propagation in the system and thus the nonlocal
contribution to the refractive index change has to be taken into account in
this case \cite{Kro}. It is worth mentioning that nonlinear media that
feature the nonlocal nonlinearity also include nonlinear ion gas \cite{Sut},
thermal nonlinear liquid \cite{Dreis}, quadratic nonlinear media \cite%
{Nikolov}, dipolar Bose-Einstein condensate \cite{Gries}, and
photorefractive crystal \cite{Segev,Saffman}.

It has been demonstrated that the propagation dynamics of beams and their
localization is significantly influenced by nonlocality \cite{Bang}. In this
respect, important results have revealed that the nonlocal nonlinearity can
suppress the modulational instability of plane waves \cite{W1,W2} and
support novel soliton states such as ring vortex solitons \cite{Bri,Kart},
gap solitons \cite{Ras}, multipole solitons \cite{Rot}, spiraling solitons 
\cite{Sku,Buc}, soliton clusters \cite{Buc1}, and incoherent solitons \cite%
{Kro2}. Although soliton structures have been extensively studied in
nonlocal Kerr-type media \cite{Kro,Dre,Kart1,Ge,Kong1,Kong2}, their
investigation in nonlocal non-Kerr systems has not been widespread. Some
significant results have, however, been obtained, with previous theoretical
studies considering nonlocality of nonlinear response and saturation \cite%
{Tsoy}. Specifically, Tsoy studied the soliton solutions in an implicit form
which propagate through a weakly nonlocal medium with cubic-quintic
nonlinearity, and derived explicit solutions in bright and dark solitons in
particular case \cite{Tsoy}.

To the best of our knowledge, investigations discussing the formation and
properties of periodic waves with nonlinear chirp in weakly nonlocal media
exhibiting cubic-quintic nonlinearity are not available to date. Moreover,
the control of chirped self-similar beams in a nonlocal nonlinear system
with distributed diffraction, cubic-quintic nonlinearity, weak nonlocality,
and gain or loss has been absent. The objective of the present work is to
study the existence and propagation properties of periodic and localized
waves with a nonlinear chirp in a weakly nonlocal cubic-quintic medium. Such
chirping property is of practical interest in achieving effective beam
compression or amplification. Additionally, the problem of self-similar
light beam propagation through a weakly nonlocal medium in the presence of
distributed cubic-quintic nonlinearity, diffraction, and gain (or loss) are
investigated too.

The paper is organized as follows. In Sec. II, we present the cubic-quintic
nonlinear Schr\"{o}dinger equation (NLSE) with weak nonlocality describing
optical beam propagation in a nonlocal medium with a saturation of the
nonlinear response. We also present here the nonlinear equation that governs
the evolution of the light beam intensity in the system and the general
traveling-wave solutions of the mentioned equation. In Sec. III, we present
results of novel chirped periodic wave solutions of the model equation and
the nonlinear chirp accompanying these nonlinear waveforms. Considering the
long-wave limit of the analytically determined periodic solutions, we find
chirped solitary beam solutions which include gray, bright and dark solitons
in Sec. IV. In Sec. V, the similarity transformation method is employed to
construct exact self-similar periodic and localized wave solutions of the
generalized NLS model with varied coefficients governing the beam evolution
in presence of the inhomogeneities of nonlocal media. We also determine here
the self-similar variables and constraints satisfied by the distributed
coefficients in the inhomogeneous NLS model. We further investigate the
propagation dynamics of the obtained self-similar localized waves (or
\textquotedblleft similaritons\textquotedblright ) in a specified soliton
control system. Finally, we give some conclusions in Sec. VI.

\section{Model of NLSE  and chirped traveling waves}

The paraxial wave equation governing one-dimensional beam propagation in a
nonlinear medium is given by \cite{Kro}:

\begin{equation}
i\frac{\partial \psi }{\partial z}+\frac{1}{2}\frac{\partial ^{2}\psi }{%
\partial x^{2}}+\Delta n\left( I\right) \psi =0,  \label{1}
\end{equation}

\noindent where $\psi \left( x,z\right) $\ is the envelope of the
electromagnetic field, $x$\ is transverse variable, and $z$\ is the
longitudinal variable representing propagation distance. Here $\Delta n$\
and $I\left( x,z\right) =\left\vert \psi \left( x,z\right) \right\vert ^{2}$
are\ the refractive index change and light intensity, respectively. In the
case of nonlocal nonlinear media, $\Delta n\left( I\right) $ can be written
in general form as \cite{K1, Kro,Tsoy}%
\begin{equation}
\Delta n\left( I\right) =\int_{-\infty }^{+\infty }R(x^{\prime
}-x)F(I(x^{\prime },z))dx^{\prime },  \label{2}
\end{equation}

\noindent where $R(x)$\ is the response function of the nonlocal medium,
which is a real symmetric function while $F\left( I\right) $\ is the
intensity-dependent function.

For weakly nonlocal media with cubic-quintic nonlinearity, one finds \cite%
{Tsoy} $\Delta n\left( I\right) =\gamma I+\sigma I^{2}+\mu \partial
_{x}^{2}I $\ with $\mu $\ being the nonlocality parameter which is defined
through $\mu =\frac{1}{2}\int_{-\infty }^{+\infty }x^{2}R(x)dx$. The
accordingly equation governing the beam propagation in such nonlinear media
is given by the NLS equation with weak nonlocality presented in \cite{Tsoy}.
For our studies, this NLS equation model is expressed as%
\begin{equation}
i\frac{\partial \psi }{\partial z}+\frac{\beta }{2}\frac{\partial ^{2}\psi }{%
\partial x^{2}}+\gamma \left\vert \psi \right\vert ^{2}\psi +\sigma
\left\vert \psi \right\vert ^{4}\psi +\mu \psi \frac{\partial ^{2}\left\vert
\psi \right\vert ^{2}}{\partial x^{2}}=0,  \label{3}
\end{equation}%
where $\gamma ,$\ $\sigma $\ and $\mu $\ \ are real parameters related to
the cubic nonlinearity, quintic nonlinearity, and weak nonlocality,
respectively, while the coefficient $\beta $ accounts for the diffraction in
the transverse plane.

In the absence of quintic nonlinearity (i.e., $\sigma =0$), Eq. (\ref{3})
becomes the modified NLS equation which applies to the description of
optical beam propagation in nonlocal nonlinear Kerr-type media \cite{Kro}.
Moreover, in the limit of vanishing weak nonlocality (i.e., $\mu =0$), Eq. (%
\ref{3}) is reduced to the cubic-quintic NLS equation which governs the
evolution of the optical beam in a local medium exhibiting third- and
fifth-order nonlinearities. As previously mentioned, the soliton solutions
in an implicit form of the model (\ref{3}) with $\beta =1,$ which are
expressed in terms of the elliptic integrals have been presented in Ref. 
\cite{Tsoy}. But here we are concerned with explicit periodic wave and
soliton solutions which are characterized by a nonlinear chirp. Our results
introduce for the first time an efficient transformation which allows the
derivation of periodic and localized solutions of Eq. (\ref{3}) in an
explicit form.

To obtain exact traveling wave solutions to the cubic-quintic NLS equation
with weak nonlocality (\ref{3}), we assume a solution given by the
expression,%
\begin{equation}
\psi \left( x,z\right) =u(\xi )\exp [i\left( \kappa z-\delta x\right) +i\phi
(\xi )],  \label{4}
\end{equation}%
where both $u(\xi )$\ and $\phi (\xi )$\ are real functions of the traveling
coordinate $\xi =x-vz$, with $v$\ being the transverse velocity of the wave.
Also $\kappa $\ and $\delta $\ represent the propagation constant and
frequency shift, respectively. Substitution of Eq. (\ref{4}) into Eq. (\ref%
{3}) and separation of the real and imaginary parts of the equation yields
the following two coupled ordinary differential equations,%
\begin{equation}
\frac{\beta }{2}\left( u\frac{d^{2}\phi }{d\xi ^{2}}+2\frac{d\phi }{d\xi }%
\frac{du}{d\xi }\right) -\left( v+\beta \delta \right) \frac{du}{d\xi }=0,
\label{5}
\end{equation}%
\begin{equation}
\frac{\beta }{2}\frac{d^{2}u}{d\xi ^{2}}-\left( \kappa -v\frac{d\phi }{d\xi }%
\right) u-\frac{\beta }{2}\left( \frac{d\phi }{d\xi }-\delta \right)
^{2}u+\gamma u^{3}+\sigma u^{5}+2\mu \left[ u\left( \frac{du}{d\xi }\right)
^{2}+u^{2}\frac{d^{2}u}{d\xi ^{2}}\right] =0.  \label{6}
\end{equation}%
The multiplication of Eq. (\ref{5}) by the function $u(\xi )$\ and
integration of the resulting equation leads to the following equation,%
\begin{equation}
\beta u^{2}\frac{d\phi }{d\xi }-(v+\beta \delta )u^{2}=J,  \label{7}
\end{equation}%
where $J$ is the integration constant. Then Eq. (\ref{7}) yields the
following expression,%
\begin{equation}
\frac{d\phi }{d\xi }=\delta +\frac{v}{\beta }+\frac{J}{\beta u^{2}(\xi )}.
\label{8}
\end{equation}%
The accompanying chirp $\Delta \omega $ defined as $\Delta \omega =-\partial %
\left[ \kappa z-\delta x+\phi (x)\right] /\partial x$ is given by%
\begin{equation}
\Delta \omega =-\frac{v}{\beta }-\frac{J}{\beta u^{2}(\xi )}.  \label{9}
\end{equation}%
Further insertion of the result (\ref{8}) into (\ref{6}) gives to the
following nonlinear ordinary differential equation, 
\begin{equation}
\frac{d^{2}u}{d\xi ^{2}}+a\left[ u\left( \frac{du}{d\xi }\right) ^{2}+u^{2}%
\frac{d^{2}u}{d\xi ^{2}}\right] +bu+cu^{3}+du^{5}-\frac{J^{2}}{\beta
^{2}u^{3}}=0,  \label{10}
\end{equation}%
where the parameters $a,$\ $b,$\ $c$\ and $d$\ are defined by%
\begin{equation}
a=\frac{4\mu }{\beta },~~~~b=\frac{v^{2}+2\beta \left( \delta v-\kappa
\right) }{\beta ^{2}},~~~~c=\frac{2\gamma }{\beta },~~~~d=\frac{2\sigma }{%
\beta }.  \label{11}
\end{equation}

\noindent Multiplying Eq. (\ref{10}) by $du/d\xi $\ and integrating the
resultant equation, we obtain 
\begin{equation}
\left( 1+au^{2}\right) \left( \frac{du}{d\xi }\right) ^{2}+bu^{2}+\frac{1}{2}%
cu^{4}+\frac{1}{3}du^{6}+\frac{J^{2}}{\beta ^{2}u^{2}}+C=0,  \label{12}
\end{equation}%
where $C$ is another integration constant. We define new function $f(\xi
)=u^{2}(\xi )$ which transforms Eq. (\ref{12}) to the following ordinary
differential equation,

\begin{equation}
\left( 1+af\right) \left( \frac{df}{d\xi }\right) ^{2}=\nu_{0}+
\nu_{1}f+\nu_{2}f^{2}+\nu_{3}f^{3}+\nu_{4}f^{4},  \label{13}
\end{equation}%
where 
\begin{equation}
\nu_{0}=-\frac{4J^{2}}{\beta ^{2}},\quad \nu_{1}=-4C,\quad \nu_{2}=-4b,\quad
\nu_{3}=-2c,\quad \nu_{4}=-\frac{4d}{3}.  \label{14}
\end{equation}

Equation (\ref{13}) presents one of the main results of our analysis,
describing the evolution of beam intensity in a weakly nonlocal medium with
cubic-quintic nonlinearity. This equation allow us to know whether the
traveling-wave solutions exist in the nonlocal medium and in what parametric
conditions they are formed. In general, this nonlinear differential equation
with coexisting\ $f\left( df/d\xi \right) ^{2}$ and $f^{4}$ terms is
difficult to handle analytically. However, by introducing a special
transformation in this paper, Eq. (\ref{13}) is solved analytically to
obtain a rich variety of nonlinear waveforms for the model (\ref{3}). Such a
transformation, to our knowledge not used before testifies about the novelty
of the solutions obtained.

Incorporating these results back into Eq. (\ref{4}), we find that general
form of traveling wave solutions to the cubic-quintic NLS equation with weak
nonlocality (\ref{3}) is%
\begin{equation}
\psi \left( x,z\right) =\pm \sqrt{f(\xi )}\exp [i\left( \kappa z-\delta
x\right) +i\phi (\xi )],  \label{15}
\end{equation}

\noindent where $f(\xi )$\ satisfies Eq. (\ref{13}) while $\phi (\xi )$ can
be evaluated explicitly using Eq. (\ref{8})\ as%
\begin{equation}
\phi \left( \xi \right) =\left( \delta +\frac{v}{\beta }\right) (\xi -\eta )+%
\frac{J}{\beta }\int_{\eta }^{\xi }\frac{1}{u^{2}(\xi )}d\xi +\phi _{0},
\label{16}
\end{equation}%
with $\phi _{0}$\ being the initial phase and $\eta $ is an arbitrary
constant.

This result shows that the phase modification $\phi (\xi )$ involves a
nonlinear contribution that is inversely proportional to light beam
intensity $\left\vert \psi \left( x,z\right) \right\vert ^{2}=\left\vert
u(\xi )\right\vert ^{2}$. Interestingly, the nontrivial nature of the phase
leads to the formation of chirped beams in the system. In particular, when $%
J=0$, the phase $\phi (\xi )$ in (\ref{16}) can be reduced to a simple
linear form as $\phi \left( \xi \right) =\left( \delta +v\beta ^{-1}\right)
(\xi -\eta )+\phi _{0}.\ $In what follows, we are interested in periodic and
solitary pulse solutions to Eq. (\ref{13}) in the most general case when $%
J\neq 0,$ which describe nonlinearly chirped structures to the model (\ref{3}).

\section{Periodic wave solutions}

As previously noted, the general solutions to Eq. (\ref{3}) which are
implicit and are expressed in terms of the elliptic integrals have been
found in \cite{Tsoy}. In this section, we introduce an efficient
transformation that enables one to obtain explicit solutions of the full
underlying cubic-quintic NLS equation with weak nonlocality (\ref{3}).
Interestingly, exact chirped periodic solutions are found in the presence of
all material parameters for the first time.

Applying the transformation (\ref{1a}) to Eq. (\ref{13}), one obtains a
modified nonlinear differential equation of the form [see Appendix A]: 
\begin{equation}
\left( \frac{df}{d\xi }\right) ^{2}=\alpha _{0}+\alpha _{1}f+\alpha
_{2}f^{2}+\alpha _{3}f^{3},  \label{17}
\end{equation}%
with the new coefficients $\alpha _{i}$ $(i=1,2,3)$ that are found in the
Appendix A as 
\begin{equation}
\alpha _{0}=-\frac{4J^{2}}{\beta ^{2}},\quad \alpha _{1}=-\frac{4b}{a}+\frac{%
2c}{a^{2}}-\frac{4d}{3a^{3}},  \label{18}
\end{equation}%
\begin{equation}
\alpha _{2}=-\frac{2c}{a}+\frac{4d}{3a^{2}},\quad \alpha _{3}=-\frac{4d}{3a}.
\label{19}
\end{equation}%
Thus the coefficient $\alpha _{0}$ in Eq. (\ref{17}) is a free parameter
because $J$ is the integration constant. Note that in Eq. (\ref{13}) there
are two free coefficients as $\nu _{0}$ and $\nu _{1}$ because $J$ and $C$
are two independent integration constants. However, Eq. (\ref{17}) has only
one free parameter $\alpha _{0}$. Thus one can use different parameter $%
\alpha _{0}$ for different solutions of Eq. (\ref{17}).

We now introduce a new function $y(\xi )$ as 
\begin{equation}
y(\xi )=-\alpha _{3}f(\xi ),\quad f(\xi )=u^{2}(\xi ).  \label{20}
\end{equation}%
Thus the equation for function $y(\xi )$ is 
\begin{equation}
\left( \frac{dy}{d\xi }\right) ^{2}=c_{0}+c_{1}y+c_{2}y^{2}-y^{3},
\label{21}
\end{equation}%
where the coefficients $c_{n}$ are given by 
\begin{equation}
c_{0}=\alpha _{0}\alpha _{3}^{2},\quad c_{1}=-\alpha _{1}\alpha _{3},\quad
c_{2}=\alpha _{2}.  \label{22}
\end{equation}%
We also introduce the polynomial $P(y)=c_{0}+c_{1}y+c_{2}y^{2}-y^{3}$ which
is given by the right side of Eq. (\ref{21}). The roots of polynomial $P(y)$
are given by equation, 
\begin{equation}
y^{3}-\alpha _{2}y^{2}+\alpha _{1}\alpha _{3}y-c_{0}=0,  \label{23}
\end{equation}%
where the coefficient $c_{0}=\alpha _{0}\alpha _{3}^{2}$ is a free parameter
because $\alpha _{0}=-4J^{2}/\beta ^{2}$ and $J$ is integration constant.

The periodic bounded solution of Eq. (\ref{21}) defined in the interval $%
y_{2}\leq y(\xi) \leq y_{3}$ is 
\begin{equation}
y(\xi)=y_{2}+(y_{3}-y_{2})\mathrm{cn}^{2}(w(\xi-\eta),k),  \label{24}
\end{equation}%
where the roots are real and ordered ($y_{1}<y_{2}<y_{3}$), and $\mathrm{cn}%
(z,k)$ is elliptic Jacobi function. The parameters $w$ and $k$ in this
solution are 
\begin{equation}
w=\frac{1}{2}\sqrt{y_{3}-y_{1}},\quad k=\sqrt{\frac{y_{3}-y_{2}}{y_{3}-y_{1}}%
} ,  \label{25}
\end{equation}%
where $0<k<1$. It is shown in the Appendix B that ordered real roots are 
\begin{equation}
y_{1}=\frac{\alpha_{2}}{3}+\frac{4(k^{2}-2)w^{2}}{3},\quad y_{2}=\frac{%
\alpha_{2}}{3}+\frac{4(1-2k^{2})w^{2}}{3},  \label{26}
\end{equation}%
\begin{equation}
y_{3}=\frac{\alpha_{2}}{3}+\frac{4(1+k^{2})w^{2}}{3}.  \label{27}
\end{equation}
The parameter $w$ in these equations is given by 
\begin{equation}
w=\frac{1}{2}\left(\frac{\alpha_{2}^{2}-3\alpha_{1}\alpha_{3}}{k^{4}-k^{2}+1}%
\right)^{1/4},  \label{28}
\end{equation}%
where it is assumed that $\alpha_{2}^{2}-3\alpha_{1}\alpha_{3}>0$. It is
shown in the Appendix B that the integration constant $J$ is fixed by
relation, 
\begin{equation}
J^{2}=-\frac{\beta^{2}y_{1}y_{2}y_{3}}{4\alpha_{3}^{2}}.  \label{29}
\end{equation}%
Hence, in general case the roots $y_{n}$ must satisfy the condition $%
y_{1}y_{2}y_{3}\leq 0$.

\textit{1. Family of chirped periodic bounded waves with $J\neq 0$.}

The amplitude in Eq. (\ref{4}) is $u(\xi)=\pm\sqrt{-y(\xi)/\alpha_{3}}$
which lead to a family of periodic bounded solutions of Eq. (\ref{3}) (with $%
0<k<1$) as 
\begin{equation}
\psi \left( x,z\right) =\pm\left[A +B\mathrm{cn}^{2}(w(\xi-\eta),k) \right]
^{1/2}\exp [i\left( \kappa z-\delta x\right) +i\phi (\xi )],  \label{30}
\end{equation}%
where the parameters $A=-y_{2}/\alpha_{3}$ and $B=-(y_{3}-y_{2}/\alpha_{3}$
are 
\begin{equation}
A=\frac{1}{3\alpha_{3}}[-\alpha_{2}+4(2k^{2}-1)w^{2}],\quad B=-\frac{%
4k^{2}w^{2}}{\alpha_{3}}.  \label{31}
\end{equation}
The parameter $w$ in this solution is given in Eq. (\ref{28}). It follows
from this solution that the parameters $A$ and $B$ should satisfy the
conditions $A> 0$ and $A+B\geq 0$. Moreover, the conditions $\alpha_{2}^{2}-3%
\alpha_{1}\alpha_{3}>0$ and $y_{1}y_{2}y_{3}< 0$ are also necessary for this
family of periodic solutions with integration constant $J\neq 0$.

We note that parameter $w$ is given by Eq. (\ref{28}), however one can
consider $w$ as independent variable parameter in Eqs. (\ref{30}) and (\ref%
{31}) because the parameter $\alpha_{1}$ depends on $b$ which is variable parameter (see Eqs. (\ref{11}) and (\ref{18})). In this case using Eq. (\ref{28}) and relation $\alpha_{1}=-4b/a-\alpha_{2}/a$ we obtain the equation for the wave number $\kappa$ as 
\begin{equation}
\kappa =v\delta +\frac{v^{2}}{2\beta }+\frac{\beta \alpha _{2}}{8}+\frac{%
a\beta \alpha _{2}^{2}}{ 24\alpha _{3}}-\frac{2a\beta w^{4}}{3\alpha _{3}}%
(k^{4}-k^{2}+1).  \label{32}
\end{equation}
This equation means that the variable parameters $w$, $v$, $\delta$ and $%
\kappa$ are connected by this relation for fixed parameter $k$ of Jacoby
elliptic function $\mathrm{cn}(\zeta,k)$. We present below three particular
cases of periodic solutions with the integration constant $J=0$.

\textit{2. Bounded periodic dn-waves for the condition $y_{1}=0$.}

The solution in Eq. (\ref{30}) for $y_{1}=0$ ($J=0$) and $0<k<1$ reduces to
the periodic waves, 
\begin{equation}
\psi \left( x,z\right) =\pm\Lambda\mathrm{dn}(w(\xi-\eta),k) \exp
(i\theta(x,z)),  \label{33}
\end{equation}%
where the parameters $\Lambda$ and $w$ are 
\begin{equation}
\Lambda=\sqrt{\frac{-\alpha_{2}}{\alpha_{3}(2-k^{2})}},\quad w=\frac{1}{2}%
\sqrt{\frac{\alpha_{2}}{2-k^{2}}}.  \label{34}
\end{equation}
The necessary conditions for this solution are $\alpha_{2}>0$ and $%
\alpha_{3}<0$. The phase $\theta(x,z)$ in this periodic solution is 
\begin{equation}
\theta(x,z)=\left( \kappa-v\delta-\frac{v^{2}}{\beta}\right)z +\frac{v}{\beta%
}x+\theta_{0},  \label{35}
\end{equation}
where $\theta_{0}=\phi_{0}-\eta(\delta+v/\beta)$. In this case ( $y_{1}=0$)
the parameter $w$ is fixed by Eq. (\ref{34}) and hence Eq. (\ref{32}) has
the form, 
\begin{equation}
\kappa =v\delta +\frac{v^{2}}{2\beta }+\frac{\beta \alpha _{2}}{8}+\frac{%
a\beta \alpha_{2}^{2}(1-k^{2})}{8\alpha_{3}(2-k^{2})^{2}}.  \label{36}
\end{equation}

\textit{3. Bounded periodic cn-waves for the condition $y_{2}=0$.}

The solution in Eq. (\ref{30}) for $y_{2}=0$ ($J=0$) and $0<k<1$ reduces to
the periodic waves, 
\begin{equation}
\psi \left( x,z\right) =\pm\Lambda\mathrm{cn}(w(\xi-\eta),k) \exp
(i\theta(x,z)),  \label{37}
\end{equation}%
where the parameters $\Lambda$ and $w$ are 
\begin{equation}
\Lambda=\sqrt{\frac{-\alpha_{2}k^{2}}{\alpha_{3}(2k^{2}-1)}},\quad w=\frac{1%
}{2}\sqrt{\frac{\alpha_{2}}{2k^{2}-1}}.  \label{38}
\end{equation}
The necessary conditions for this solution are $\alpha_{2}>0$ and $%
\alpha_{3}<0$ for $1/\sqrt{2}<k<1$; and $\alpha_{2}<0$ and $\alpha_{3}<0$
for $0<k<1/\sqrt{2}$. The phase $\theta(x,z)$ in this solution is given by
Eq. (\ref{35}). In this case ( $y_{2}=0$) the parameter $w$ is fixed by Eq. (%
\ref{38}) and hence Eq. (\ref{32}) has the form, 
\begin{equation}
\kappa =v\delta +\frac{v^{2}}{2\beta }+\frac{\beta \alpha _{2}}{8}-\frac{%
a\beta \alpha_{2}^{2}k^{2}(1-k^{2})}{8\alpha_{3}(2k^{2}-1)^{2}}.  \label{39}
\end{equation}

\textit{4. Bounded periodic sn-waves for the condition $y_{3}=0$.}

The solution in Eq. (\ref{30}) for $y_{3}=0$ ($J=0$) and $0<k<1$ reduces to
the periodic waves, 
\begin{equation}
\psi \left( x,z\right) =\pm\Lambda\mathrm{sn}(w(\xi-\eta),k) \exp
(i\theta(x,z)),  \label{40}
\end{equation}%
where the parameters $\Lambda$ and $w$ are 
\begin{equation}
\Lambda=\sqrt{\frac{-\alpha_{2}k^{2}}{\alpha_{3}(1+k^{2})}},\quad w=\frac{1}{%
2}\sqrt{\frac{-\alpha_{2}}{1+k^{2}}}.  \label{41}
\end{equation}
The necessary conditions for this solution are $\alpha_{2}<0$ and $%
\alpha_{3}>0$. The phase $\theta(x,z)$ in this solution is given by Eq. (\ref%
{35}). In this case ($y_{3}=0$) the parameter $w$ is fixed by Eq. (\ref{41})
and hence Eq. (\ref{32}) has the form, 
\begin{equation}
\kappa =v\delta +\frac{v^{2}}{2\beta }+\frac{\beta \alpha _{2}}{8}+\frac{%
a\beta \alpha_{2}^{2}k^{2}}{8\alpha_{3}(1+k^{2})^{2}}.  \label{42}
\end{equation}
We emphases that in Eqs. (\ref{36}), (\ref{39}) and (\ref{42}) there are
three variable parameters: $\delta$, $v$ and $\kappa$ for fixed parameter $k$%
. This means that one can fix any two of these variable parameters and then
the third variable parameter follows from Eqs. (\ref{36}), (\ref{39}) and (%
\ref{42}) respectively. For an example, if the variable parameters $\delta$
and $\kappa$ are fixed then the above equations yield the parameter $v$.

\section{Solitary wave solutions}

In this section, we present various nonlinearly chirped solitary wave
solutions of the cubic-quintic NLS equation with weak nonlocality (\ref{3}).
At first we consider the solution in Eq. (\ref{30}) in the limiting case $%
k=1 $ for integration constant $J\neq 0$. In this case the periodic bounded
solution in Eq. (\ref{30}) reduces to solitary wave as 
\begin{equation}
\psi \left( x,z\right) =\pm \left[ A+B\mathrm{sech}^{2}(w_{0}(\xi -\eta ))%
\right] ^{1/2}\exp [i\left( \kappa z-\delta x\right) +i\phi (\xi )],
\label{43}
\end{equation}%
where the inverse width is $w_{0}=\frac{1}{2}(\alpha _{2}^{2}-3\alpha
_{1}\alpha _{3})^{1/4}$. It is assumed here that the condition $\alpha
_{2}^{2}-3\alpha _{1}\alpha _{3}>0$ is satisfied. The parameters $A$ and $B$
are 
\begin{equation}
A=\frac{1}{3\alpha _{3}}(4w_{0}^{2}-\alpha _{2}),\quad B=-\frac{4w_{0}^{2}}{%
\alpha _{3}}.  \label{44}
\end{equation}%
Note that in the case with $k=1$ the roots are 
\begin{equation}
y_{1}=\frac{\alpha _{2}}{3}-\frac{4w_{0}^{2}}{3},\quad y_{2}=\frac{\alpha
_{2}}{3}-\frac{4w_{0}^{2}}{3},\quad y_{3}=\frac{\alpha _{2}}{3}+\frac{%
8w_{0}^{2}}{3},  \label{45}
\end{equation}%
where $y_{1}=y_{2}$. The equation for variable parameters $w_{0}$, $v$, $%
\delta $ and $\kappa $ for this case ($k=1$) are connected by relation, 
\begin{equation}
\kappa =v\delta +\frac{v^{2}}{2\beta }+\frac{\beta \alpha _{2}}{8}+\frac{%
a\beta \alpha _{2}^{2}}{24\alpha _{3}}-\frac{2a\beta w_{0}^{4}}{3\alpha _{3}}%
.  \label{46}
\end{equation}%
We consider here the case with integration constant $J\neq 0$, and hence Eq.
(\ref{29}) yields the condition $y_{3}<0$ because $y_{1}=y_{2}$ for $k=1$.
It follows from Eq. (\ref{45}) that the condition $y_{3}<0$ is satisfied for 
$\alpha _{2}<-8w_{0}^{2}$, and hence we have $\alpha _{2}<0$. Moreover, in this case ($J\neq 0$) we have $y_{n}\neq 0$, and hence $4w_{0}^{2}-\alpha _{2}\neq 0$
and $A\neq 0$. The parameter $A$ in solution (\ref{43}) should be positive which is possible only in the case with $\alpha _{3}>0$ because we have $\alpha _{2}<0$ (see Eq. (\ref{44})). 
It also follows from Eq. (\ref{44}) that $A+B=-(8w_{0}^{2}+\alpha _{2})/3\alpha _{3}>0$ because $y_{3}<0$ and $\alpha _{3}>0$. Thus we have found that the conditions $A>0$, $B<0$ and $A+B>0$ are satisfied for the solution given in Eq. (\ref{43}). These conditions lead to gray soliton solution which is a dark soliton with nonzero minimum intensity.

\textit{5. Chirped gray soliton solution.}

The solution given in Eq. (\ref{43}) with $J\neq 0$ can also be written in
the equivalent form as 
\begin{equation}
\psi \left( x,z\right) =\pm\left[\Lambda +D\mathrm{tanh}^{2}(w_{0}(\xi-%
\eta)) \right] ^{1/2}\exp [i\left( \kappa z-\delta x\right) +i\phi (\xi )],
\label{47}
\end{equation}%
\begin{figure}[h]
\includegraphics[width=1.25\textwidth]{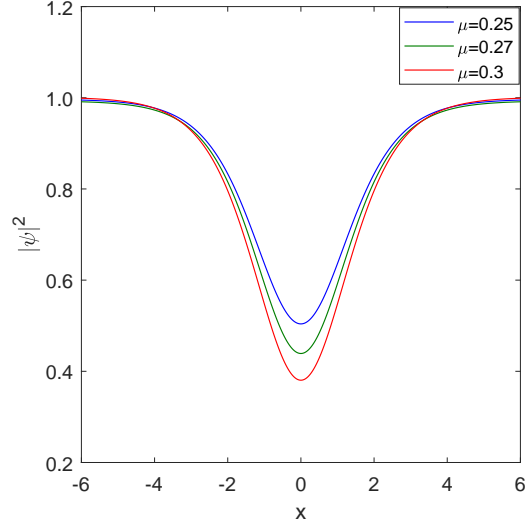}
\caption{Intensity profiles of the chirped gray solitary wave
solution (\ref{47})) for different values of the nonlocality parameter $\mu :$ $\mu=0.25,$ $\mu =0.27,$ $\mu =0.30$. Other parameters are $\beta =1,$ $\sigma=-1,$ $\gamma =1,$ $v=0.1,$ $\delta =2.81,$ $\kappa =0.12,$ $\eta =0$.}
\label{FIG.1.}
\end{figure}
where the inverse width of the gray soliton is $w_{0}=\frac{1}{2}%
(\alpha_{2}^{2}-3\alpha_{1}\alpha_{3})^{1/4}$ with the condition $%
\alpha_{2}^{2}>3\alpha_{1}\alpha_{3}$. The parameters $\Lambda=A+B$ and $%
D=-B $ are given by 
\begin{equation}
\Lambda=-\frac{1}{3\alpha_{3}}(\alpha_{2}+8w_{0}^{2}),\quad D=\frac{
4w_{0}^{2}}{\alpha_{3}}.  \label{48}
\end{equation}%
We have shown that this solution occur only in the case when $%
\alpha_{2}<-8w_{0}^{2}$ and $\alpha _{3}>0$. These conditions also lead to
inequalities $\Lambda >0$ and $D>0$. Hence, the minimum intensity of the
pulse is $I=\Lambda\neq 0$ which typically for gray solitons. The equation
for variable parameters $w_{0}$, $v$, $\delta$ and $\kappa$ for this case ($%
k=1$) are connected by Eq. (\ref{46}).

We note that the condition $\Lambda D>0$ is satisfied for gray soliton
solution. In this case the phase $\phi (\xi )$ in Eq. (\ref{47}) follows by
Eq. (\ref{16}) as 
\begin{eqnarray}
\phi \left( \xi \right) &=&-\frac{JD}{2\beta Rw_{0}}\arcsin \left( \frac{%
(\Lambda +D)+(\Lambda -D)\cosh (2w_{0}(\xi -\eta ))}{(\Lambda -D)+(\Lambda
+D)\cosh (2w_{0}(\xi -\eta ))}\right)  \notag \\
&&+\left( \delta +\frac{v}{\beta }+\frac{J}{\beta (\Lambda +D)}\right) (\xi
-\eta )+\phi _{0},~~~~~~~~  \label{49}
\end{eqnarray}%
where $R=\sqrt{\Lambda D}(\Lambda +D)$.

Typical intensity profiles of the chirped solitary wave (\ref{47}) at $z=0$
are shown in Fig. 1 for different degrees of nonlocality $\mu $ as: $\mu
=0.25,$ $\mu =0.27,$ $\mu =0.30$. The parameter values used are $\beta =1,$ $%
\sigma =-1,$ $\gamma =1,$ $v=0.1,$ $\delta =2.81,$ $\kappa =0.12,$ $\eta =0.$
%From this figure, one can clearly see that for this case of strong
%nonlocality, all chirped gray solitary waves exhibit the same width as they
%propagate through the weakly nonlocal medium due to the presence of
%cubic-quintic nonlinearities. Similar results have been also found for dark
%solitons propagating in weakly nonlocal media with Kerr nonlinearity \cite{Kro}. 
An interesting observation in this figure is that the minimum intensity of chirped
gray solitary wave decreases continuously with the increasing of the degree
of nonlocality. We also find that the background intensity of the chirped
gray solitary waves remains the same for different values of the nonlocality
parameter $\mu $. 
%We can conclude that in the limit of strong nonlocality,
%the width and background of propagating gray solitary waves being
%independent of the degree of nonlocality.

\textit{6. Bright soliton solution.}

The solutions in Eqs. (\ref{33}) and (\ref{37}) for limiting case $k=1$ and $%
J=0$ reduce to chirped bright soliton solution, 
\begin{equation}
\psi \left( x,z\right) =\pm \sqrt{-\frac{\alpha _{2}}{\alpha _{3}}}\mathrm{%
sech}(\frac{\sqrt{\alpha _{2}}}{2}(\xi -\eta ))\exp (i\theta (x,z)),
\label{50}
\end{equation}%
\begin{figure}[h]
\includegraphics[width=1.25\textwidth]{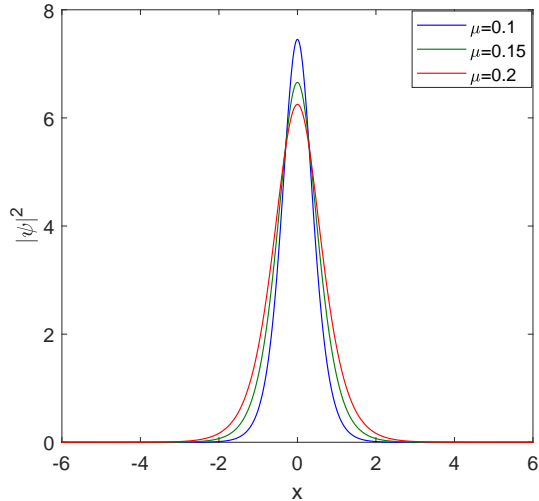}
\caption{ Intensity profiles of the chirped bright solitary wave
solution (\ref{50}) for different values of the nonlocality parameter $\mu :$ $\mu
=0.10,$ $\mu =0.15,$ $\mu =0.20$. Other parameters are $\beta =1,$ $\gamma
=-1,$ $\sigma =0.3,$ $v=0.1,$ $\delta =2.81,$ $\kappa =0.12,$ $\eta =0$.}
\label{FIG.2.}
\end{figure}
where the necessary conditions for this solution are $\alpha _{2}>0$ and $%
\alpha _{3}<0$. In this bright soliton solution the phase $\theta (x,z)$ is
given by Eq. (\ref{35}). Note that we have $c_{0}=y_{2}^{2}y_{3}$ which
leads to relation $c_{0}=\alpha _{0}\alpha _{3}^{2}=0$ because $y_{2}=0$.
This yields $\alpha _{0}=0$ and integration constant $J=0$. Moreover, Eqs. (\ref{36}) and (\ref{39}) lead to equation for variable parameters $\delta $
and $\kappa $ and $v$ as 
\begin{equation}
\kappa =v\delta +\frac{v^{2}}{2\beta }+\frac{\beta \alpha _{2}}{8}.
\label{51}
\end{equation}%
Note that in this case the inverse width is fixed as $w_{0}=\sqrt{\alpha _{2}%
}/2$.

Figure 2 depicts the intensity profiles of the chirped bright solitary wave (%
\ref{50}) for different values of the nonlocality parameter $\mu .$ It can
be observed that with the increasing of nonlocality, the intensity of the
solitary wave gradually decreases, while the width increases leading to
broadening of the light beam if the parameter $\mu $ is further increased.

\textit{7. Dark soliton solution.}

The solution in Eq. (\ref{40}) for limiting case $k=1$ and $J=0$ reduces to
chirped dark soliton solution, 
\begin{equation}
\psi (x,z)=\pm \sqrt{-\frac{\alpha _{2}}{2\alpha _{3}}}\mathrm{tanh}(\sqrt{-%
\frac{\alpha _{2}}{8}}(\xi -\eta ))\exp (i\theta (x,z)), 
\label{52}
\end{equation}%
\begin{figure}[h]
\includegraphics[width=1.25\textwidth]{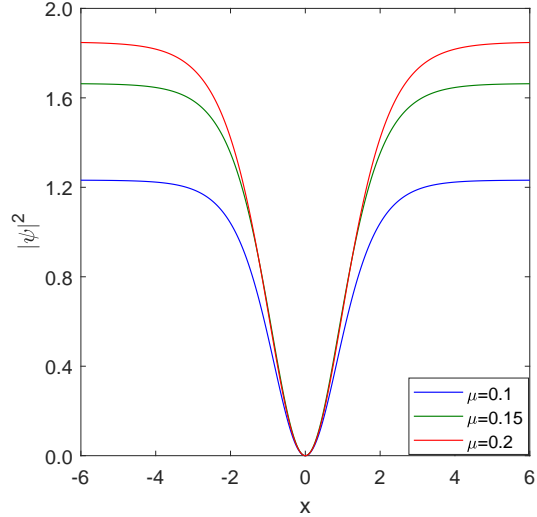}
\caption{Intensity profiles of of the chirped dark solitary wave
solution (\ref{52}) for different values of the nonlocality parameter $\mu :$ $\mu=0.10,$ $\mu =0.15,$ $\mu =0.20$. Other parameters are $\beta =-1,$ $\gamma=1,$ $\sigma =-0.3,$ $v=0.1,$ $\delta =2.81,$ $\kappa =0.12,$ $\eta =0$.}
\label{FIG.3.}
\end{figure}
where the necessary conditions for this solution are $\alpha _{2}<0$ and $%
\alpha _{3}>0$. In this dark soliton solution the phase $\theta (x,z)$ is
given by Eq. (\ref{35}). Note that in this case we have $%
c_{0}=y_{2}^{2}y_{3} $ where $y_{3}=0$, and hence $c_{0}=\alpha _{0}\alpha
_{3}^{2}=0$ which yields $\alpha _{0}=0$ and integration constant $J=0$.
Moreover, Eq. (\ref{42}) leads to equation for variable parameters $\delta $
and $\kappa $ and $v$ as 
\begin{equation}
\kappa =v\delta +\frac{v^{2}}{2\beta }+\frac{\beta \alpha _{2}}{8}+\frac{%
a\beta \alpha _{2}^{2}}{32\alpha _{3}}.  \label{53}
\end{equation}%
In this case the inverse width is fixed as $w_{0}=\sqrt{-\alpha _{2}/8}$. We
note that in Eqs. (\ref{51}) and (\ref{53}) there are three variable
parameters: $\delta $, $v$ and $\kappa $. This means that one can fix any
two of these variable parameters and then the third parameter follows from
above equations respectively.

The intensity profile of the chirped dark solitary wave (\ref{52}) for
different values of $\mu $ is displayed in Fig. 3. 
One can see that unlike chirped gray solitary waves, the background intensity of the chirped
dark solitary waves increase as the degree of nonlocality grows.
%the chirped bright solitary waves, the width of the chirped dark solitary
%wave decreases as the degree of nonlocality increases.

\section{Similarity transformation of generalized CQNLS equation}

In a real optical material including nonlocal nonlinear media, the physical
parameters vary along with the propagation of light beams due to the
presence of the inhomogeneities of media. The inclusion of the distributed
coefficients into the NLS equations is currently an effective way to reflect
the inhomogeneous effects of the optical beams \cite{LWang}. In what
follows, we analyze the beam propagation phenomena in a realistic weakly
nonlocal cubic-quintic medium exhibiting varied physical parameters. An
accurate description of the beam evolution in such inhomogeneous system can
be achieved by means of variation with respect to the propagation distance
of all the material parameters in Eq. (\ref{3}), resulting in the
generalized NLS model with distributed coefficients: 
\begin{equation}
i\frac{\partial \Phi }{\partial s}+\frac{D(s)}{2}\frac{\partial ^{2}\Phi }{%
\partial \chi ^{2}}+R_{1}(s)\left\vert \Phi \right\vert ^{2}\Phi
+R_{2}(s)\left\vert \Phi \right\vert ^{4}\Phi +N(s)\Phi \frac{\partial
^{2}(\left\vert \Phi \right\vert ^{2})}{\partial \chi ^{2}}=iG(s)\Phi ,
\label{54}
\end{equation}%
where $D(s),$ $R_{1}(s)$ and $R_{2}(s)$ represent the variable diffraction,
cubic and quintic nonlinearity coefficients, respectively. Parameter $N(s)$
denotes the weak nonlocality coefficient, while $G(s)$ represents the loss ($%
G(s)<0$) or gain ($G(s)>0$) coefficient.

In order to connect solutions of Eq. (\ref{54}) with those of Eq. (\ref{3}),
we will use the transformation \cite{Dai1,Dai4,Dai3} as 
\begin{equation}
\Phi (s,\chi )=\rho (s)\psi \left[ x(s,\chi ),z(s)\right] \exp \left[
i\varphi (s,\chi )\right] ,  \label{55}
\end{equation}%
where $\rho (s),$ $z(s),$ $x(s,\chi ),$ and $\varphi (s,\chi )$ are real
functions to be determined. Substituting Eq. (\ref{55}) into Eq. (\ref{54})
leads to Eq. (\ref{3}), but we must have the following set of parametric
equations: 
\begin{equation}
2\rho _{s}+D\rho \varphi _{\chi \chi }-2G\rho =0,~~~~x_{s}+Dx_{\chi }\varphi
_{\chi }=0,  \label{56}
\end{equation}%
\begin{equation}
2\phi _{s}+D\left( \varphi _{\chi }\right) ^{2}=0,\quad D\left( x_{\chi
}\right) ^{2}-\beta z_{s}=0,  \label{57}
\end{equation}%
\begin{equation}
N\rho ^{2}\left( x_{\chi }\right) ^{2}-\mu z_{s}=0,~~~~x_{\chi \chi }=0,
\label{58}
\end{equation}%
\begin{equation}
R_{1}\rho ^{2}-\gamma z_{s}=0,~~~~R_{2}\rho ^{4}-\sigma z_{s}=0,  \label{59}
\end{equation}

\noindent where subscripts denote partial differentiation. These equations
can be solved self-consistently to obtain the self-similar wave amplitude $%
\rho (s)$ and phase $\varphi (s,\chi )$:%
\begin{eqnarray}
&&\left. \rho (s)=\frac{\sqrt{\gamma D(s)}}{\sqrt{\beta R_{1}(s)}W(s)}%
,\right.  \label{60} \\
&&\left. \varphi (s,\chi )=-\frac{c_{0}\Gamma (s)}{2}\chi ^{2}-b_{0}\Gamma
(s)\chi -\frac{b_{0}^{2}}{2}\Gamma (s)d(s),\right.  \label{61}
\end{eqnarray}

\noindent together with the similarity variable $x(s,\chi )$ and effective
propagation distance $z(s)$:%
\begin{eqnarray}
&&\left. x(s,\chi )=\frac{\chi -\chi _{c}(s)}{W(s)},\right.  \label{62} \\
&&\left. z(s)=\frac{d(s)\Gamma (s)}{\beta W_{0}^{2}},\right.  \label{63}
\end{eqnarray}

\noindent where the beam width $W(s)$ and center position $\chi _{c}(s)$ are
given by

\begin{equation}
W(s)=W_{0}/\Gamma (s),\quad \chi _{c}(s)=\chi _{0}-\left( b_{0}+c_{0}\chi
_{0}\right) d(s).  \label{64}
\end{equation}

\noindent Meanwhile, the accumulted diffraction $d(s)$ and the parameter
related to the phase-front curvature wave $\Gamma (s)$ are given by

\begin{equation}
d(s)=\int_{0}^{s}D(s^{\prime })ds^{\prime },\quad \Gamma (s)=\left[
1-c_{0}d(s)\right] ^{-1}.  \label{65}
\end{equation}%
Here the parameters $\chi _{0}$, $W_{0}$, $c_{0}$ and $b_{0}$ are constants
representing the initial values of central position of beam, width, chirp,
and position of the wavefront, respectively. Further, the constraint
conditions on the management parameters depicting weak nonlocality, quintic
nonlinearity, and gain (or loss) are given by%
\begin{eqnarray}
&&\left. N(s)=\frac{\mu R_{1}(s)W^{2}(s)}{\gamma },\right.  \label{66} \\
&&\left. R_{2}(s)=\frac{\beta \sigma R_{1}^{2}(s)W^{2}(s)}{\gamma ^{2}D(s)}%
,\right.  \label{67} \\
&&\left. G(s)=\frac{1}{2}\left\{ \frac{\mathcal{W}\left[ R_{1}(s),D(s)\right]
}{R_{1}(s)D(s)}-\frac{c_{0}W_{0}D(s)}{W(s)}\right\}, \right.  \label{68}
\end{eqnarray}

\noindent with the notation for the Wronskian $\mathcal{W}\left[
R_{1}(s),D(s)\right] =R_{1}D_{s}-DR_{1s}.$

We notice that the accumulated diffraction $d(s)$ influences not only the
characteristics of the self-similar pulse such as the amplitude, width,
center position, and phase but also the effective propagation distance.

\noindent Hence, a similarity transformation between Eq. (\ref{54}) and Eq. (%
\ref{3}) can be obtained as%
\begin{equation}
\Phi (s,\chi )=\frac{\sqrt{\gamma D(s)}}{\sqrt{\beta R_{1}(s)}W(s)}\psi %
\left[ x(s,\chi ),z(s)\right] e^{i\varphi (s,\chi )},  \label{69}
\end{equation}

\noindent where the phase $\varphi (s,\chi )$ is given by Eq. (\ref{61}) and 
$\psi \left( x,z\right) $ satisfies Eq. (\ref{3}).

With this transformation [Eq. (\ref{69})], one can construct exact
self-similar solutions of the generalized NLS equation model (\ref{54}) by
using the above analytic solutions of the constant-coefficient NLS model (%
\ref{3}). Let us first construct exact self-similar periodic wave solutions
of Eq. (\ref{54}). Using the transformation (\ref{69}) with Eqs. (\ref{66})-(%
\ref{68}) and periodic bounded solution (\ref{30}) of Eq. (\ref{3}), we
obtain a self-similar cnoidal wave solution of the generalized NLS equation (%
\ref{54}) in the form%
\begin{equation}
\Phi (s,\chi )=\pm \frac{\sqrt{\gamma D(s)}}{\sqrt{\beta R_{1}(s)}W(s)}\left[
A+B\mathrm{cn}^{2}(w\zeta ,k)\right] ^{1/2}\exp \left[ i\Theta (s,\chi )%
\right] ,  \label{70}
\end{equation}%
where the traveling coordinate $\zeta $ is given by%
\begin{equation}
\zeta (s,\chi )=\frac{\Gamma (s)\left\{ \chi +\left( b_{0}+c_{0}\chi
_{0}\right) d(s)-\chi _{0}\right\} }{W_{0}}-\frac{v}{\beta W_{0}^{2}}%
d(s)\Gamma (s)-\eta ,  \label{71}
\end{equation}

\noindent and the phase of field has the form,%
\begin{equation}
\Theta (s,\chi )=-\frac{c_{0}\Gamma (s)}{2}\chi ^{2}-\left( b_{0}+\frac{%
\delta }{W_{0}}\right) \Gamma (s)\chi -\left\{ \frac{b_{0}^{2}}{2}+\frac{%
\delta \left( b_{0}+c_{0}\chi _{0}\right) }{W_{0}}-\frac{\kappa }{\beta
W_{0}^{2}}\right\} \Gamma (s)d(s)+\frac{\delta \chi _{0}}{W_{0}}\Gamma
(s)+\phi (\xi ).  \label{72}
\end{equation}

\noindent Note that the parameters $A,$ $B$ and $\kappa $ in the family of
self-similar cnoidal wave solutions (\ref{70}) are defined by Eqs. (\ref{31}%
) and (\ref{32}). While the phase structure for self-similar waves
propagating in cubic-quintic nonlinear media seem to be quadratic [see Refs. 
\cite{Sent,Dai2}], the phase of self-similar light beams in the presence of
weak nonlocality takes a more complicated form [Eq. (\ref{72})], which
involves an extra intensity-dependent phase term $\phi (\xi )$ [Eq. (\ref{16}%
)]. This implies that the derived self-similar solutions are chirped
nonlinearly which would find potential applications in light compression or
amplification.

A second family of exact self-similar periodic wave solution of the
generalized NLS equation (\ref{54}) can be obtained by inserting the
solution (\ref{33}) into the transformation (\ref{69}) as%
\begin{equation}
\Phi (s,\chi )=\pm \frac{\Lambda \sqrt{\gamma D(s)}}{\sqrt{\beta R_{1}(s)}%
W(s)}\mathrm{dn}(w\zeta ,k)\exp \left[ i\Theta (s,\chi )\right] ,  \label{73}
\end{equation}

\noindent where $\zeta $ is same as the one given by Eq. (\ref{71}) while
the phase $\Theta (s,\chi )$ takes the form,
\begin{equation}
\Theta (s,\chi )=-\frac{c_{0}\Gamma (s)}{2}\chi ^{2}-\left( b_{0}-\frac{v}{%
\beta W_{0}}\right) \Gamma (s)\chi -\left\{ \frac{b_{0}^{2}}{2}-\frac{%
v\left( b_{0}+c_{0}\chi _{0}\right) }{\beta W_{0}}-\frac{\beta (\kappa
-v\delta )-v^{2}}{\beta ^{2}W_{0}^{2}}\right\} \Gamma (s)d(s)-\frac{v\chi
_{0}}{\beta W_{0}}\Gamma (s)+\theta _{0}.  \label{74}
\end{equation}

\noindent Also $\Lambda $ and $w$ satisfy Eq. (\ref{34}) and $\kappa $ is
given by Eq. (\ref{36}).

A third family of exact self-similar periodic wave solution of the model (%
\ref{54}) can be found by inserting Eq. (\ref{37}) into the transformation (%
\ref{69}) as%
\begin{equation}
\Phi (s,\chi )=\pm \frac{\Lambda \sqrt{\gamma D(s)}}{\sqrt{\beta R_{1}(s)}%
W(s)}\mathrm{cn}(w\zeta ,k)\exp \left[ i\Theta (s,\chi )\right] ,  \label{75}
\end{equation}

\noindent where the parameters{\large \ }$\Lambda $ and $w$ are given by Eq.
(\ref{38}) and $\kappa $ is shown in Eq. (\ref{39}). Meanwhile, the variable 
$\zeta $ and phase $\Theta (s,\chi )$ take the same form as Eqs. (\ref{71})
and (\ref{74}), respectively.

Another class of exact self-similar periodic wave solution of Eq. (\ref{54})
can be determined by substituting Eq. (\ref{40}) into the transformation (%
\ref{69}) as%
\begin{equation}
\Phi (s,\chi )=\pm \frac{\Lambda \sqrt{\gamma D(s)}}{\sqrt{\beta R_{1}(s)}%
W(s)}\mathrm{sn}(w\zeta ,k)\exp \left[ i\Theta (s,\chi )\right] ,  \label{76}
\end{equation}

\noindent where $\zeta $ and $\Theta (s,\chi )$ are same as the ones given
by Eqs. (\ref{71}) and (\ref{74}), respectively. Moreover, $\Lambda $ and $w$
are given by Eq. (\ref{41}) while $\kappa $ is determined by Eq. (\ref{42}).

Next we construct the exact self-similar localized solutions of the
generalized NLS equation with distributed coefficients (\ref{54}).
Substitution of the solution (\ref{43}) into the transformation (\ref{69})
yields an exact self-similar solitary wave solution of Eq. (\ref{54}) of the
form,
\begin{equation}
\Phi (s,\chi )=\pm \frac{\sqrt{\gamma D(s)}}{\sqrt{\beta R_{1}(s)}W(s)}\left[
A+B\mathrm{sech}^{2}(w_{0}\zeta )\right] ^{1/2}\exp \left[ i\Theta (s,\chi )%
\right] ,  \label{77}
\end{equation}

\noindent where $A\ $and $B$ are defined by Eq. (\ref{44}), $\zeta $ and $%
\Theta (s,\chi )$ are shown in Eqs. (\ref{71}) and (\ref{72}) respectively,
with $\kappa $ given by Eq. (\ref{46}).

Moreover, substitution of the solution (\ref{47}) into the transformation (%
\ref{69}) leads to an exact chirped self-similar gray soliton solution of
Eq. (\ref{54}) of the form,
\begin{equation}
\Phi (s,\chi )=\pm \frac{\sqrt{\gamma D(s)}}{\sqrt{\beta R_{1}(s)}W(s)}\left[
\Lambda +D\mathrm{tanh}^{2}(w_{0}\zeta )\right] ^{1/2}\exp \left[ i\Theta
(s,\chi )\right] ,  \label{78}
\end{equation}%
where $\Lambda $ and $D$ are shown in Eq. (\ref{48}) and $\zeta $ and $%
\Theta (s,\chi )$ are given by Eqs. (\ref{71}) and (\ref{72}) respectively,
with the phase shift $\phi (\xi )$ given by the relation (\ref{49}).

It is interesting that we can construct a chirped self-similar bright-type
soliton solution for the generalized NLS equation (\ref{54}) by combining
Eqs. (\ref{50}) and (\ref{69}) as%
\begin{equation}
\Phi (s,\chi )=\pm \frac{\sqrt{\gamma D(s)}}{\sqrt{\beta R_{1}(s)}W(s)}\sqrt{%
-\frac{\alpha _{2}}{\alpha _{3}}}\mathrm{sech}(\frac{\sqrt{\alpha _{2}}}{2}%
\zeta )\exp \left[ i\Theta (s,\chi )\right] ,  \label{79}
\end{equation}

\noindent in the case where $\alpha _{3}<0$ and $\alpha _{2}>0.$ It should
be noted here that $\zeta $ and $\Theta (s,\chi )$ take the same expressions
stated in (\ref{71}) and (\ref{74}) respectively and $\kappa $ is given by
Eq. (\ref{51}).

A chirped self-similar dark-type soliton solution can be also obtained for
Eq. (\ref{54}) by using Eqs. (\ref{52}) and (\ref{69}) as%
\begin{equation}
\Phi (s,\chi )=\pm \frac{\sqrt{\gamma D(s)}}{\sqrt{\beta R_{1}(s)}W(s)}\sqrt{%
-\frac{\alpha _{2}}{2\alpha _{3}}}\mathrm{tanh}(\sqrt{-\frac{\alpha _{2}}{8}}%
\zeta )\exp \left[ i\Theta (s,\chi )\right] ,  \label{80}
\end{equation}

\noindent when $\alpha _{2}<0$ and $\alpha _{3}>0.$ Here the variable $\zeta 
$ and phase $\Theta (s,\chi )$ are same as the ones given by Eqs. (\ref{71})
and (\ref{74}) while the wave number $\kappa $ satisfies Eq. (\ref{53}).

Now we discuss how to realize the control for chirped self-similar beams
presented above. To demonstrate the controllable self-similar structures,
here we take the chirped self-similar localized solutions (\ref{78}) and (%
\ref{79}) as examples to\ discuss the dynamical behaviors of the
self-similar gray and bright solitary waves in the weakly nonlocal medium.
In this situation, we take a periodic distributed amplification system with
varying diffraction and nonlinear parameters of the form \cite{Dai2,RH,Tang}:%
\begin{equation}
D(s)=D_{0}\cos \left( gs\right) ,\quad R_{1}(s)=R_{0}\exp \left( -rs\right)
\cos \left( gs\right) ,  \label{81}
\end{equation}

\noindent where $D_{0}$ and $g$ are the parameters used to describe the
diffraction, while $R_{0}$ and $r$ are related to the nonlinearity. The
corresponding weak nonlocality, quintic nonlinearity, and gain (or loss) of
the optical medium defined by Eqs. (\ref{66}), (\ref{67}) and (\ref{68}) read%
\begin{eqnarray}
&&\left. N(s)=\frac{\mu R_{0}W_{0}^{2}\cos \left( gs\right) \exp \left(
-rs\right) }{\gamma }\left[ 1-\epsilon \sin \left( gs\right) \right]
^{2},\right.  \label{82} \\
&&\left. R_{2}(s)=\frac{\sigma \beta R_{0}^{2}W_{0}^{2}\cos \left( gs\right)
\exp \left( -2rs\right) }{\gamma ^{2}D_{0}}\left[ 1-\epsilon \sin \left(
gs\right) \right] ^{2},\right.  \label{83} \\
&&\left. G(s)=\frac{r}{2}-\frac{c_{0}D_{0}\cos \left( gs\right) }{2\left[
1-\epsilon \sin \left( gs\right) \right] },\right.  \label{84}
\end{eqnarray}%
where we introduced for brevity the parameter $\epsilon =c_{0}D_{0}/g.$
\begin{figure}[h]
\includegraphics[width=1.25\textwidth]{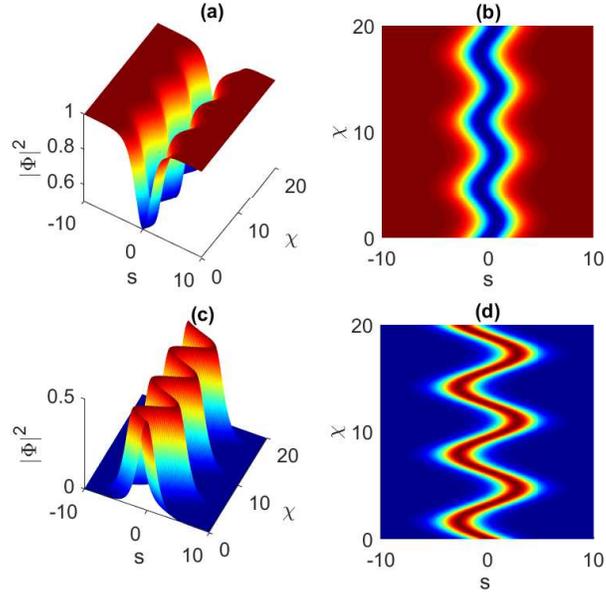}
\caption{Evolution of self-similar solitary wave solutions with
parameters $\mu =0.125,$ $g=1$, $D_{0}=R_{0}=1,$ $r=0,$ $c_{0}=0,$ $b_{0}=1,$ $v=0.9,$ $W_{0}=1,$ $\delta =1.2,$ $\kappa =1.807,$ $\chi_{0}=\eta =0$; (a)-(b) gray self-similar wave when $\beta =0.5,$ $\gamma =0.5,$ $\sigma=-0.5$; (c)-(d) bright self-similar wave when $\beta =-0.5,$ $\gamma =-0.5,$ $\sigma =0.5$.}
\label{FIG.4.}
\end{figure}
\begin{figure}[h]
\includegraphics[width=1.25\textwidth]{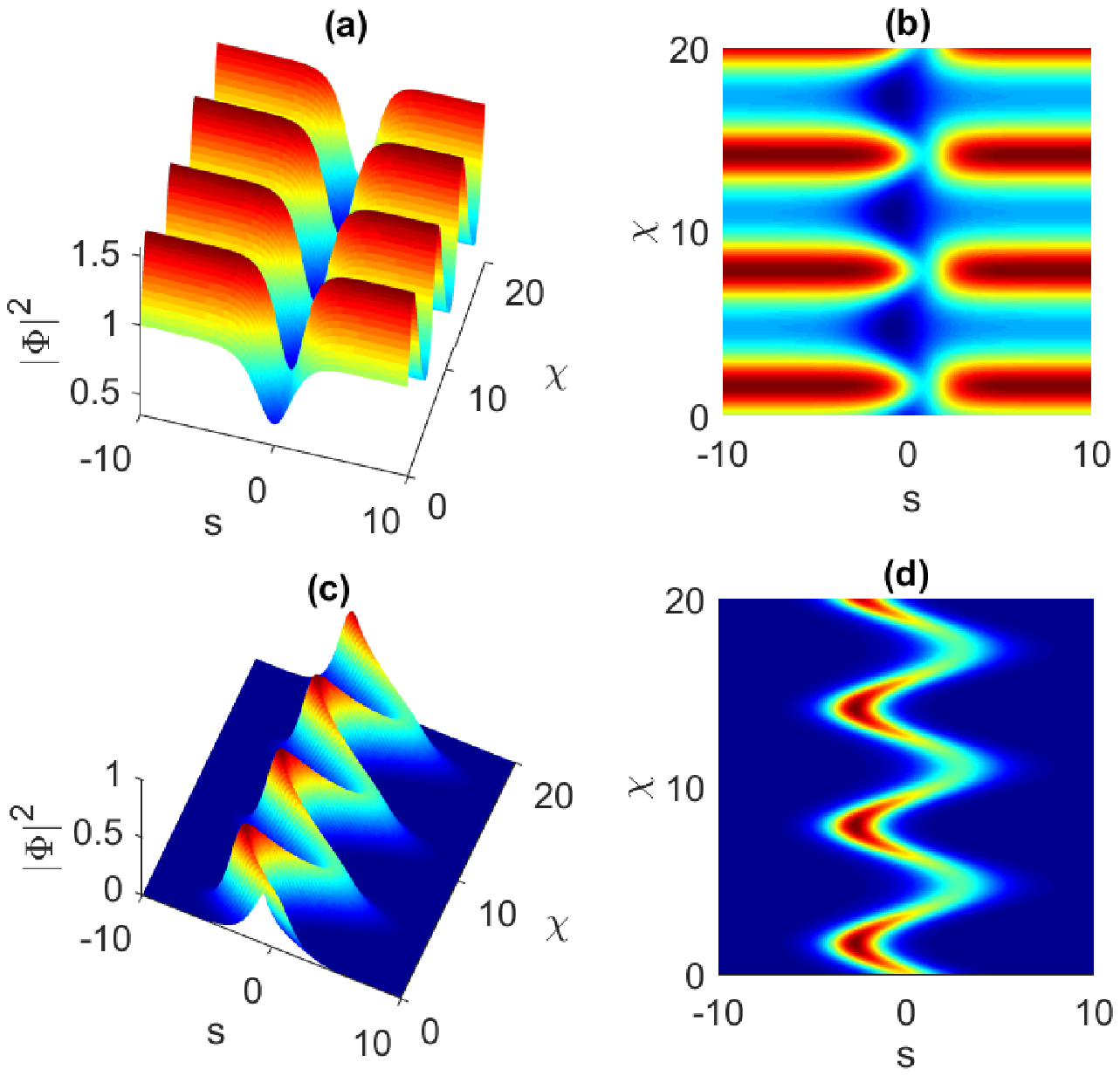}
\caption{Evolution of chirped self-similar wave solutions with
parameters $c_{0}=0.2,$ $b_{0}=1,$ $v=0.9$; (a)-(b) gray self-similar wave;
(c)-(d) bright self-similar wave. Other parameters are same as given in Fig.4.}
\label{FIG.5.}
\end{figure}
\begin{figure}[h]
\includegraphics[width=1.25\textwidth]{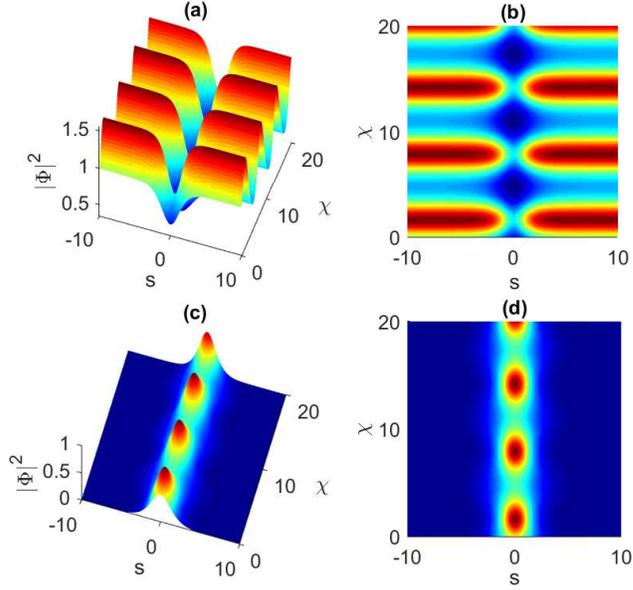}
\caption{Evolution of chirped self-similar wave solutions with
parameters $c_{0}=0.2,$ $b_{0}=0,$ $v=0$; (a)-(b) gray self-similar wave;
(c)-(d) bright self-similar wave. Other parameters are same as given in Fig.4.}
\label{FIG.6.}
\end{figure}

First, we analyze the evolutional behavior of self-similar solitary beams in
the case when the initial chirp $c_{0}=0$. In this situation, the gain
(loss) function (\ref{84}) takes a constant form $G(s)=r/2$, which
corresponds to the diffraction decreasing nonlocal medium for $r<0$. Here,
without loss of generality, we use the parameters $D_{0}=1$, $g=1$, and $%
R_{0}=1$ and choose the value of the nonlocality parameter as $\mu =0.125$.
Figure 4 shows the evolution and contour plots of the chirped self-similar
solitary wave solutions (\ref{78}) and (\ref{79}) with $D(s)$ and $R_{1}(s)$
given by Eq. (\ref{81}) for the parameters $b_{0}=1,$ $r=0,$ $v=0.9,$ $%
\delta =1.2,$ $\kappa =1.807$, $W_{0}=1,$ and $\chi _{0}=\eta =0.$ From it,
one observes a snake-like behavior of the gray and bright solitary beams as
they propagate through the weakly nonlocal medium. For such snaking-like
trajectory, the profile of the self-similar solitary waves maintains its
structural integrity in propagating along the system although its position
varies periodically. Notice that for this case, the central position as
given by Eq. (\ref{64}) reads $\chi _{c}(s)=\chi _{0}-\left[ \left(
b_{0}+c_{0}\chi _{0}\right) D_{0}/g\right] \sin \left( gs\right) ,$ thus
indicating that it oscillates periodically along distance even if the value
of initial chirp is zero (i.e., $c_{0}=0$). However, for the same case, the
width given by Eq. (\ref{64}) which obeys the relation $W(s)=W_{0}\left[
1-\epsilon \sin \left( gs\right) \right] $, becomes a constant $W(s)=W_{0}$
when $c_{0}=0.$ One should note here that the snakelike evolutional behavior
of the gray and bright structures takes place due to the presence of the
parameters $b_{0}$ and $v$ in the traveling coordinate $\zeta $ [Eq. (\ref%
{71})].

Next, we discuss the dynamical behavior of self-similar waves in the case of
practical interest $c_{0}\neq 0.$ Considering the initial chirp parameter $%
c_{0}=0.2$ and using the same values of $b_{0}$ and $v$ as in Fig. 4, one
can see that the chirped self-similar gray and bright solitary beams show a
periodical change in their intensity but their profile remain unchanged in
propagation along the nonlocal medium [Figs. 5(a)-(b)\ and 5(c)-(d)]. We can
conclude that the oscillation of the self-similar beam intensity\ here
results from the term $c_{0}d(s)$\ in the expression for the amplitude given
by Eq. (\ref{60}).\textbf{\ }Taking the same value of the initial chirp $%
c_{0}=0.2$ and for $b_{0}=0\ $and $v=0$, we see that that the chirping leads
to localization and yields the periodic emergence of solitary waves [Figs.
6(a)-(b)\ and 6(c)-(d)]. We note in passing that similar behaviors can also
be obtained in the case of the chirped self-similar dark-type soliton
solutions (\ref{80}). Hence, we can conclude that rich and significant
self-similar solitary beam dynamics can be obtained in the soliton control
system (\ref{81}) by choosing the parameters $c_{0},$ $b_{0}$ and $v$
appropriately$.$\newline

\section{Conclusion}

To conclude, we studied the formation and properties of spatial nonlinear
waves in a weakly nonlocal media with cubic-quintic nonlinearity. A wide
variety of periodic waves which are characterized by a nonlinear chirp have
been found in the system for the first time. The chirping property of such
nonlinear period waves makes them of practical importance for achieving
effective beam compression or amplification. It is remarked that these
chirped periodic structures do exist in nonlocal media in the presence of
all material parameters. The explicit chirped gray optical dip solutions as
well as chirped bright and dark soliton solutions are obtained in the
long-wave limit of the derived periodic waveforms. In addition, we have
investigated the self-similar propagation of optical beams in an
inhomogeneous weakly nonlocal media wherein the light propagation is
described by the generalized NLS equation with distributed weak nonlocality,
diffraction, cubic-quintic nonlinearities and gain or loss. Interestingly,
the wider class of periodic and localized solutions presented here possess
attractive features such as nonlinearity in beam chirp and self-similarity
in beam shape. We also discussed the applications of the obtained chirped
self-similar localized beams by considering a periodic distributed
amplification system. The results showed that the dynamical behaviors of
those nonlinearly chirped self-similar beams could be controlled by choosing
appropriate diffraction and quintic nonlinearity parameters. Future research
includes a systematic study of the stability of such nonlinearly chirped
solitary waves with respect to finite perturbations such as amplitude
perturbation, the slight violation of the parametric conditions, and random
noises.

\appendix

\section{Modified nonlinear differential equation}

In this Appendix we consider the transformation of Eq. (\ref{13}) to
modified Eq. (\ref{17}). Note that such transformation is possible because
the coefficients $\nu_{0}$ and $\nu_{1}$ in Eq. (\ref{13}) are free
parameters. These coefficients are free because $\nu_{0}$ and $\nu_{1}$
depend on integration constants $J$ and $C$. We define the coefficients $%
\alpha_{n}$ by equation, 
\begin{equation}
\sum_{n=0}^{4}\nu_{n}f^{n}(\xi)=(1+af(\xi))\sum_{n=0}^{3}\alpha_{n}f^{n}(%
\xi).  \label{1a}
\end{equation}%
Equation (\ref{1a}) yields the system of algebraic equations as 
\begin{equation}
\alpha_{0}=\nu_{0},\quad
\alpha_{1}+a\alpha_{0}=\nu_{1},\quad\alpha_{2}+a\alpha_{1}=\nu_{2},\quad
\alpha_{3}+a\alpha_{2}=\nu_{3},  \label{2a}
\end{equation}%
where the last equation is  
\begin{equation}
a\alpha_{3}=\nu_{4}.  \label{3a}
\end{equation}%
Relations (\ref{2a}) lead to the coefficients as
\begin{equation}
\alpha_{0}=\nu_{0},\quad \alpha_{1}=\nu_{1}-a\nu_{0},\quad
\alpha_{2}=\nu_{2}-a\nu_{1}+a^{2}\nu_{0},  \label{4a}
\end{equation}%
\begin{equation}
\alpha_{3}=\nu_{3}-a\nu_{2}+a^{2}\nu_{1}-a^{3}\nu_{0}.  \label{5a}
\end{equation}%
We emphases that Eqs. (\ref{3a}) and (\ref{5a}) yield the constraint for
coefficients $\nu_{n}$ as 
\begin{equation}
\nu_{4}=a\nu_{3}-a^{2}\nu_{2}+a^{3}\nu_{1}-a^{4}\nu_{0}.  \label{6a}
\end{equation}
Equations (\ref{4a}), (\ref{5a}) and the constraint (\ref{6a}) lead to the
following coefficients, 
\begin{equation}
\alpha_{0}=\nu_{0},\quad \alpha_{1}=\frac{\nu_{2}}{a}-\frac{\nu_{3}}{a^{2}}+%
\frac{\nu_{4}}{a^{3}},\quad \alpha_{2}=\frac{\nu_{3}}{a}-\frac{\nu_{4}}{a^{2}%
},\quad \alpha_{3}=\frac{\nu_{4}}{a}.  \label{7a}
\end{equation}%
Note that one can assume that the constraint (\ref{6a}) is satisfied because
the coefficients $\nu_{0}$ and $\nu_{1}$ are free parameters. These
coefficients are free because they depend on integration constants $J$ and $%
C $ (see Eq. (\ref{14})). Moreover, one can consider the coefficient $%
\nu_{0} $ as a free parameter and $\nu_{1}$ is fixed by (\ref{6a}) for an
arbitrary given $\nu_{0}$. The substitution of Eq. (\ref{1a}) to Eq. (\ref%
{13}) leads to modified nonlinear differential equation, 
\begin{equation}
\left( \frac{df}{d\xi }\right) ^{2}=\alpha_{0}+
\alpha_{1}f+\alpha_{2}f^{2}+\alpha_{3}f^{3},  \label{8a}
\end{equation}%
where the coefficients $\alpha_{n}$ are given by Eq. (\ref{7a}). In this
equation the coefficient $\alpha_{0}$ is a free parameter because $%
\alpha_{0}=\nu_{0}$. We emphases that in Eq. (\ref{13}) there are two free
coefficients as $\nu_{0}$ and $\nu_{1}$. However, the modified Eq. (\ref{8a}) has one free coefficient $\alpha_{0}$.

\section{Ordered real roots of polynomial}

In this Appendix we present three ordered real and different roots $%
y_{1}<y_{2}<y_{3}$ in Eq. (\ref{23}). The polynomial $%
P(y)=c_{0}+c_{1}y+c_{2}y^{2}-y^{3}$ can be written as 
\begin{equation}
c_{0}+c_{1}y+c_{2}y^{2}-y^{3}=-\prod_{n=1}^{3}(y-y_{n}),  \label{1b}
\end{equation}%
where $y_{n}$ are the roots in Eq. (\ref{23}). Eq. (\ref{1b}) yields the
relations, 
\begin{equation}
c_{1}=-y_{2}y_{3}-y_{1}y_{2}-y_{1}y_{3},\quad c_{2}=y_{1}+y_{2}+y_{3},
\label{2b}
\end{equation}%
\begin{equation}
c_{0}=y_{1}y_{2}y_{3},\quad c_{0}=\alpha_{0}\alpha_{3}^{2},  \label{3b}
\end{equation}%
where $c_{1}=-\alpha_{1}\alpha_{3}$ and $c_{2}=\alpha_{2}$. Moreover, Eq. (%
\ref{25}) yields the relation, 
\begin{equation}
y_{2}=k^{2}y_{1}+(1-k^{2})y_{3}.  \label{4b}
\end{equation}%
Eqs. (\ref{2b}) and (\ref{4b}) lead to the following ordered real and
different roots $y_{n}$ of the polynomial $P(y)$, 
\begin{equation}
y_{1}=\frac{\alpha_{2}}{3}+\frac{(k^{2}-2)}{3}\sqrt{\frac{%
\alpha_{2}^{2}-3\alpha_{1}\alpha_{3}}{k^{4}-k^{2}+1}},  \label{5b}
\end{equation}%
\begin{equation}
y_{2}=\frac{\alpha_{2}}{3}+\frac{(1-2k^{2})}{3}\sqrt{\frac{%
\alpha_{2}^{2}-3\alpha_{1}\alpha_{3}}{k^{4}-k^{2}+1}},  \label{6b}
\end{equation}%
\begin{equation}
y_{3}=\frac{\alpha_{2}}{3}+\frac{(1+k^{2})}{3}\sqrt{\frac{%
\alpha_{2}^{2}-3\alpha_{1}\alpha_{3}}{k^{4}-k^{2}+1}}.  \label{7b}
\end{equation}%
Note that $k^{4}-k^{2}+1>0$ for the arbitrary values of parameter $k$. One
can show that under conditions $\alpha_{2}^{2}-3\alpha_{1}\alpha_{3}\geq 0$
and $0<k<1$ these three roots are real, different and ordered as $%
y_{1}<y_{2}<y_{3}$.

It follows from Eq. (\ref{25}) that the parameter $w$ is 
\begin{equation}
w=\frac{1}{2}\sqrt{y_{3}-y_{1}}=\frac{1}{2}\left(\frac{\alpha_{2}^{2}-3%
\alpha_{1}\alpha_{3}}{k^{4}-k^{2}+1}\right)^{1/4}.  \label{8b}
\end{equation}%
Thus the parameters $A$ and $B$ in Eq. (\ref{29}) are 
\begin{equation}
A=\frac{1}{3\alpha_{3}}[-\alpha_{2}+4(2k^{2}-1)w^{2}],\quad B=-\frac{%
4k^{2}w^{2}}{\alpha_{3}}.  \label{9b}
\end{equation}%
We emphasis that the constraint in Eq. (\ref{3b}) has the form, 
\begin{equation}
J^{2}= -\frac{\beta^{2}y_{1}y_{2}y_{3}}{4\alpha_{3}^{2}},  \label{10b}
\end{equation}%
where $J$ is integration constant. Hence, this constraint is satisfied for
the integration constant $J$ given in Eq. (\ref{10b}). However, in this case
the roots $y_{n}$ must satisfy to the inequality $y_{1}y_{2}y_{3}\leq 0$.

\end{document}